\documentclass[prb,aps,twocolumn,superscriptaddress,showpacs]{revtex4}

\usepackage{graphicx}

\begin{document}

\title{Strong influence of the diffuse component on the lattice dynamics in Pb(Mg$_{1/3}$Nb$_{2/3}$)O$_{3}$}

\author{C. Stock}
\affiliation{Department of Physics, University of Toronto, Ontario, Canada
M5S 1A7}

\author{H. Luo}
\affiliation{Shanghai Institute of Ceramics, Chinese Academy of Sciences, Shanghai, China, 201800}

\author{D. Viehland}
\affiliation{Department of Materials Science and Engineering, Virginia Tech., Blacksburg, Virginia, 24061}

\author{J. F. Li}
\affiliation{Department of Materials Science and Engineering, Virginia Tech., Blacksburg, Virginia, 24061}

\author{I. Swainson}
\affiliation{National Research Council, Chalk River, Ontario, Canada, K0J
1J0}

\author{R. J. Birgeneau}
\affiliation{Department of Physics, University of Toronto, Ontario, Canada
M5S 1A7}

\author{G. Shirane}
\affiliation{Physics Department, Brookhaven National Laboratory, Upton,
New York 11973}

\date{\today}

\begin{abstract}

The temperature and zone dependence of the lattice dynamics in Pb(Mg$_{1/3}$Nb$_{2/3}$)O$_{3}$ is characterized using neutron inelastic scattering.  A strong correlation between the diffuse and phonon scattering is measured.  The lattice dynamics in Brillouin zones where the diffuse scattering is strong is observed to display qualitatively different behavior than that in zones where the diffuse scattering is weak.  In the (220) and (200) zones, where there is a weak diffuse component, the dynamics are well described by coupled harmonic oscillators.  Compared with SrTiO$_{3}$, the coupling is weak and isotropic, resulting in only a small transfer of spectral weight from one mode to another.  A comparison of the scattering in these zones to that in the (110) zone, where a strong diffuse component is present, reveals a strong coupling of the diffuse (or central) component to the acoustic mode.  We speculate that the coupling to the central peak is the reason for several recent conflicting interpretations of the lattice dynamics based on data from zones with a strong diffuse component.

\end{abstract}

\pacs{77.80.-e, 61.10.Nz, 77.84.Dy}

\maketitle

\section{Introduction}

    The relaxor ferroelectrics, with generic chemical formula Pb\textit{B}O$_{3}$, with the \textit{B} site disordered, have generated intensive research recently due to their promising applications as piezoelectric devices.~\cite{Ye98:81}  Despite much work on these materials there is no consensus on the origin of their interesting properties.  Pb(Mg$_{1/3}$Nb$_{2/3}$)O$_{3}$ (PMN) and Pb(Zn$_{1/3}$Nb$_{2/3}$)O$_{3}$ (PZN) are prototypical relaxor systems which display a diffuse transition with a broad and frequency dependent dielectric response, but no well-defined structural transition without the application of an electric field.  The nature of the diffuse transition and the low-temperature ground state is important to the complete understanding of these materials and has been the focus of much recent research and debate.

    Early studies of the refractive index by Burns and Docal (Ref. \onlinecite{Burns83:48}) found that local regions of ferroelectric order were formed in a paraelectric background at a high-temperature, denoted as the Burns temperature T$_{d}$.  The connection of this high-temperature ``transition'' to the diffuse scattering (as observed by thermal neutrons) was made by Vakhrushev \textit{et al.} and Hirota \textit{et al.} who observed that the diffuse scattering (around $\hbar \omega$=0 ) appeared at nearly the same temperature as T$_{d}$.~\cite{Vak89:90,Hirota02:65}  A central peak appearing at a temperature $\sim$ T$_{d}$ is one of the basic features of the random field model of Ref. \onlinecite{Stock03:xx}.  Recent work using cold-neutron elastic scattering (with much improved energy resolution) by Xu \textit{et al.}, Hiraka \textit{et al.}, and Gvasaliya \textit{et al.} has characterized the diffuse component in detail.~\cite{Xu03:xx,Hiraka04:xx,Gvasaliya04:49}

    Pioneering work on the lattice dynamics was originally conducted by Naberezhnov \textit{et al.}~\cite{Nab99:11} Based on this, and further experiments around the (300) position, a new soft ``quasi-optic'' mode was proposed to explain the inconsistency between the structure factors measured from the diffuse and the transverse-optic (TO) mode.~\cite{Vak02:xx}  An alternate model was suggested by Gehring \textit {et al.} and Wakimoto \textit{et al.}~\cite{Gehring01:87,Wakimoto02:66, Wakimoto02:65}  They suggested that the TO mode was the soft-mode and were able to describe the lattice dynamics around the (300) and (200) positions consistently by a coupling between the TA and TO modes.  Gehring \textit{et al.} also observed a large dampening of the TO mode for $q$ near the zone centre and attributed this to the formation of polar nanoregions.  A different model was proposed by Hlinka \textit{et al.} who were able to describe the lattice dynamics, including the anomalous dampening of the TO mode, in PMN qualitatively well using a standard mode-coupling scheme, without the need for polar nanoregions.~\cite{Hlinka03:91}

    The discrepancies in the interpretation of the lattice dynamics has been based solely on data taken in a (300) type zone, where a strong diffuse component is present.  In all analyses of the phonons in PMN, there has been little consideration of the fact that the coupling may be anisotropic and also the feature that the (300) zone has a strong diffuse peak with an inelastic component.  An example of anisotropic mode-coupling has been observed in SrTiO$_{3}$ by Yamada and Shirane.~\cite{Yamada69:26}  They found strong coupling between the acoustic and optic mode for phonons propagating along the [100] direction (measured near (200)), but little coupling was observed for phonons propagating along [110] (measured near (220)).~\cite{Yamada69:26}  This is illustrated in Fig. \ref{figure1} which plots constant-Q scans taken near (200) in SrTiO$_{3}$.  At low-temperatures (4.5 K) a nearly complete transfer of spectral weight from the TA mode to the TO mode is observed, an indication of strong coupling.

    In this paper we describe neutron inelastic results aimed at elucidating the effects of mode-coupling (including any anisotropy) and the effect of the diffuse component on the phonons.  Since all discrepancies to-date have been based on data obtained in Brillouin zones with a strong diffuse component (in particular (300)), we have focused our work in the (200) and (220) zones where there is no observable diffuse scattering.  To study the effects of the diffuse scattering we contrast the results around (220) to preliminary data taken near the (110) zone, where a significant diffuse contribution has been measured around $\hbar \omega$=0.

\section{Experiment}

\begin{figure}[t]
\includegraphics[width=7cm] {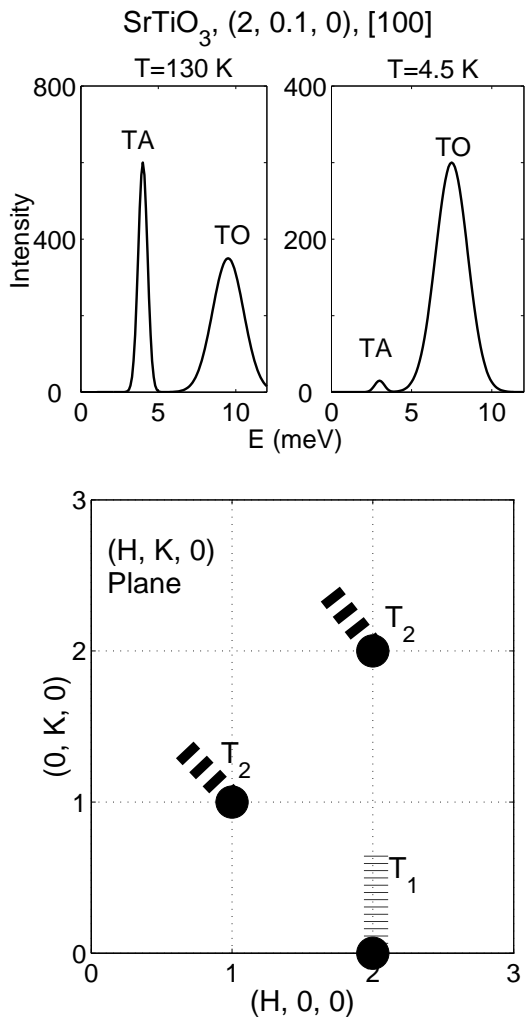}
\caption{\label{figure1} The upper panel schematically plots the effects of coupling as observed in SrTiO$_{3}$ for phonons along the [100] direction as measured in Ref. \onlinecite{Yamada69:26}.  The lower panel displays the (HK0) scattering plane and illustrates the difference between the T$_{1}$ and T$_{2}$ phonons.  The dashed lines represent the regions in momentum space where the two modes were respectively studied.}
\end{figure}

    Neutron inelastic scattering experiments were conducted at the C5 spectrometer located in the NRU reactor at Chalk River Laboratories. Constant-Q scans were made by fixing the final energy (E$_{f}$) to 14.6 meV and varying the incident energy (E$_{i}$).  A variable focusing graphite (002) monochromator and a flat graphite (002) analyzer were used with the horizontal collimation fixed at 12$'$-33$'$-S$'$-29$'$-72$'$.  A graphite filter was used on the scattered side to filter out higher order neutrons and a liquid nitrogen cooled sapphire filter was used before the monochromator to reduce the fast neutron background.  All inelastic data have been corrected for higher-order contamination of the incident beam monitor as described in detail elsewhere.~\cite{Shirane:book}

    The 9.3 cc crystal was grown by the modified Bridgeman technique previously described.~\cite{Luo00:39, Luo03:xx}  The sample was mounted in a cryofurnace such that reflections of the form (HK0) were in the scattering plane.  The room temperature lattice constant was measured to be \textit{a}=4.04 \AA.  To characterize further the sample we have measured the temperature dependence of the zone-centre transverse optic frequency and have found that it behaves identically to that previously measured on much smaller samples.  As well, when plotted as a function of reduced temperature $t=T/T_{c}$, the zone-center TO frequency agrees with that found Pb(Zn$_{1/3}$Nb$_{2/3}$)O$_{3}$.~\cite{Stock03:xx}

    We have focused our experiments in the (200) and (220) Brillouin zones where there is no observable diffuse component.  The diffuse scattering (or sometimes referred to as the central mode) has been observed to be very weak in these two zones and therefore its influence on the TA and TO modes is expected to be minimal.  This approach differs from that in previous studies of the lattice dynamics which have addressed the problem by comparing the dynamics in the (200) zone (where the diffuse component is weak) to that of the (300) zone (where the diffuse scattering is strong).

    As illustrated in Fig. \ref{figure1} we have focused our measurements on both the T$_{1}$ and T$_{2}$ phonons by conducting constant-Q scans in the (200) and the (220) zones.  In the (200) zone, we have studied transverse acoustic T$_{1}$ phonons with the propagation vector along the [010] direction and polarization along [100].  For scans near (220), we have investigated T$_{2}$ phonons with propagation vector along [1$\overline{1}$0] and polarization along [110].  Therefore, any anisotropy in the lattice dynamics can be studied by comparing these two zones without the influence of a strong diffuse component.

\begin{figure}[t]
\includegraphics[width=7cm] {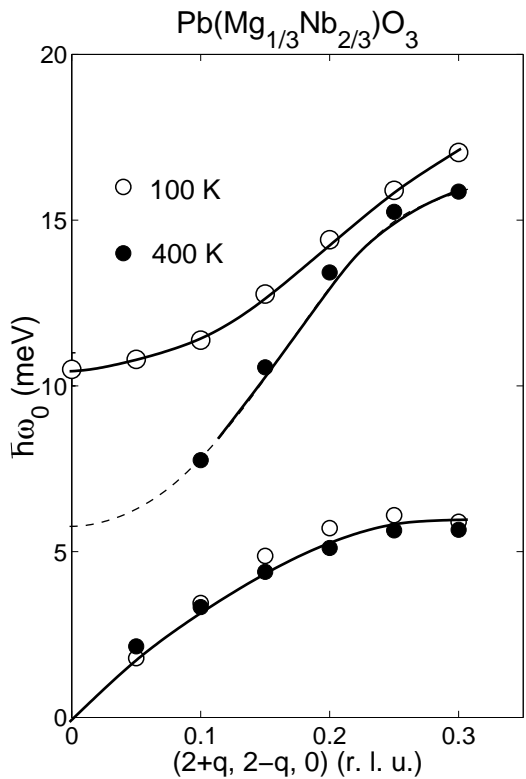}
\caption{\label{figure2} The dispersion of the TA and TO modes is plotted at 400 K and 100 K.  The solid lines are the results of fits to the dispersion discussed in the text to extract the zone centre frequency of the TO mode.  The 400 K dotted line indicates where the TO mode becomes overdamped and hence difficult to observe in a constant-Q scan.}
\end{figure}

    To extract a linewidth, amplitude, and frequency we have fit the data to two uncoupled harmonic oscillators described by Lorentzian lineshapes as discussed in the Appendix (Eqn. \ref{SHO}).  The acoustic dispersion was approximated to be linear, $\omega_{\circ}({\bf{q}})=c |\bf{q}|$, where $c$ is the phonon velocity.  For optic modes, we have set the dispersion to have the form

\begin{eqnarray}
\label{optic}
\omega_{\circ}({\bf{q}})^{2}=\Omega_{0}^{2}+\alpha |{\bf{q}}|^2,
\end{eqnarray}

\noindent where $\Omega_{\circ}$ is the soft-mode frequency at $q=0$ and $\alpha$ is a constant for a given $\bf{q}$ direction.~\cite{Shirane70:2,Axe70:1}  For temperatures around T$_{c}$ we have used this dispersion to extract the zone center frequency $\Omega_{\circ}$ by extrapolating from higher $q$, where the optic phonon is underdamped as discussed in the next section.

    For our coupled mode analysis we have used the same expressions presented in Ref. \onlinecite{Currat89:40}, \onlinecite{Harada71:4}, Refs. \onlinecite{Wakimoto02:66}, and described in the Appendix (Eqn. \ref{coupled}). We have followed the coupling analyses of Refs. \onlinecite{Wakimoto02:66}, \onlinecite{Gehring01:63}, and Ref. \onlinecite{Hlinka03:91} by taking the coupling constant to be real.

    To extract the temperature dependence of the structure factors we have used the formula for the intensity predicted by harmonic theory.  The neutron cross-section as integrated over $\hbar \omega$ for a single phonon at $\omega_{0}$ is expressed as,

\begin{eqnarray}
\label{Harmonic_structure} {{d\sigma} \over {d\Omega}}\propto {\hbar \over {2\omega_{0}}}\mid F(q) \mid^{2} Q^{2} [n(\omega_{0})+1],
\end{eqnarray}

\noindent where $\mid F(q) \mid$ is the structure factor at a particular momentum transfer $q$ and $(n+1)$ is the Bose factor.  As measured in SrTiO$_{3}$, and as discussed in the next section, the temperature dependence of the structure factors is a sensitive measure of the coupling strength between the TA and TO modes.

\section{(220) and (200) Brillouin Zones}

\begin{figure}[t]
\includegraphics[width=7cm] {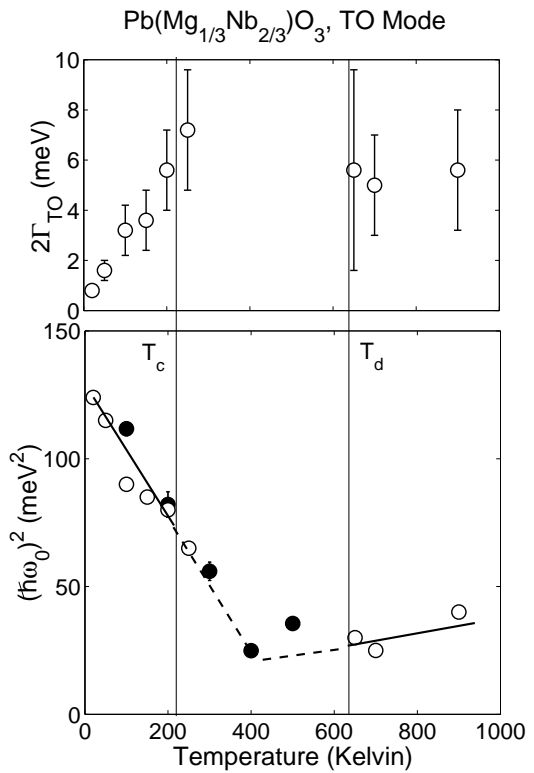}
\caption{\label{figure3} The TO zone centre frequency squared and the FWHM are plotted as a function of temperature.  The open circles are from Ref. \onlinecite{Wakimoto02:65} (around (200)) and the filled circles are data obtained in this study by extrapolating the measured dispersion relation to the (220) zone centre from higher $q$ (as discussed in the text).}
\end{figure}

\begin{figure}[t]
\includegraphics[width=8cm] {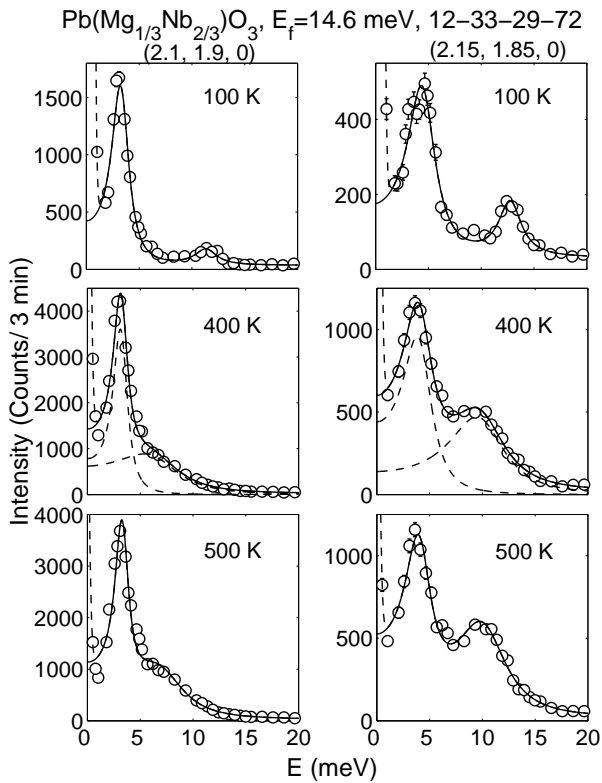}
\caption{\label{figure4} Constant-Q scans at 100 K, 400 K, and 100 K, are plotted for \textit{q} near (220).  The solid lines are the results of fits to uncoupled harmonic oscillators.  The parameters extracted from these fits were used to calculate the structure factors displayed in Fig. 5.}
\end{figure}

    As stated above, to investigate the TA and TO phonons we have conducted constant-Q scans near the (220) and (200) positions (see Fig. \ref{figure4}).  The dispersion for both the TA and TO modes is summarized in Fig. \ref{figure2} and the temperature dependence of the full-width and the frequency position are plotted in Fig. \ref{figure3}.  Fig. \ref{figure4} shows that at low temperatures the TO mode is underdamped while at higher temperatures the TO mode becomes heavily damped near the zone centre and is barely observable at 400 K at $q$=0.1 r.l.u.  Despite the fact that the TO mode becomes highly damped at $q$=0.1, scans at $q$=0.15 and larger values of $q$ show underdamped TO modes.   This indicates that the anomalous damping of the TO mode around T$_{c}$ is strictly limited to values of $q$ near the zone centre.

    To extract the zone centre frequency we have measured the TO frequency position at large $q$, where an underdamped TO peak is observed, and extrapolated to lower $q$ using the TO frequency dispersion described in the previous section.  The results of the extrapolated zone centre frequency positions are plotted in Fig. \ref{figure3} and are compared to those measured in Ref. \onlinecite{Wakimoto02:66} using constant-Q scans at the zone centre.  The measured zone frequencies are also in good agreement with those obtained from infrared spectroscopy.~\cite{Bovtun03:NATO}  The agreement between the two sets of measurements is quite good and validates the extrapolation technique used in this work to extract the zone centre frequency.  Concomitant with the softening of the TO zone centre frequency, we also observe an increase of the TO damping for modes near the zone centre.  A similar broadening of the TO mode around T$_{c}$ has also been observed in PZN.\cite{Stock03:xx}

    A qualitative method of characterizing the effects of mode coupling is to study the temperature dependence of the integrated intensities as well as the frequency positions.  In the case of SrTiO$_{3}$ a large shift of spectral weight is observed to occur for phonons propagating along [100], but not [110] therefore indicating anisotropic mode coupling.~\cite{Yamada69:26}  In KTaO$_{3}$, strong coupling is also observed along the [100] direction as deduced by a large shift in the TA frequency with TO frequency.~\cite{Axe70:1}  Theories showing that that the observed frequency positions depends strongly on coupling have been described in Ref. \onlinecite{Yamada81:50} and Ref. \onlinecite{Harbeke70:8}

    Motivated by these previous studies we have characterized the effects of coupling by studying the phonon frequencies and the integrated intensities as functions of temperature.  The temperature dependence of the frequency positions is illustrated in Fig. \ref{figure2} where we show the dispersion of the TA and TO modes in the (220) zone at 400 K and 100 K.  Despite a significant softening of the TO mode near the zone centre, the TA mode does not shift in frequency significantly.  This result is confirmed by the constant-Q scans displayed in Fig. \ref{figure4} which show little change in the acoustic frequency despite a large variation in the optic frequency.  Based on the measured temperature dependences of the TA and TO frequencies, we infer that there are no strong coupling effects.

    To investigate the coupling near (220) and (200) more quantitatively, we have measured the structure factors from constant-Q scans for various values of $q$ ranging from $q$=0.05 r.l.u to 0.3 r.l.u, examples of which are displayed in Fig. \ref{figure4}.  To extract the peak positions, amplitudes, and linewidths of the TA and TO modes we have used the uncoupled harmonic oscillator lineshape convolved with the resolution function as described in the Appendix.  A nonconvoluted Gaussian centered at $\hbar\omega$=0 together with a constant term were used to describe the background.

\begin{figure}[t]
\includegraphics[width=8cm] {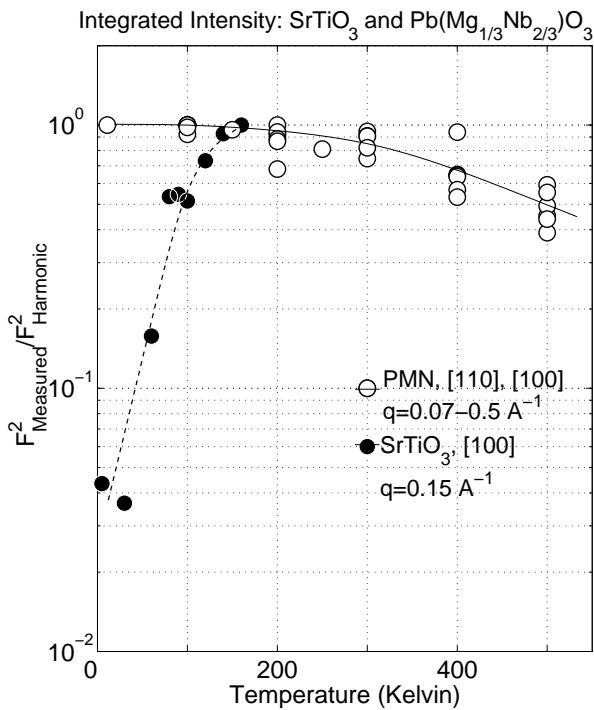}
\caption{\label{figure6} The measured acoustic structure factors, on a logarithmic scale,  as a function of temperature is plotted for PMN (open circles and solid line) and SrTiO$_{3}$ (filled circles and dashed line).  SrTiO$_{3}$ shows a large change in the structure factors for phonons propagating along [100], indicative of mode coupling.  PMN, in comparison, shows only a mild change in integrated intensity. We have taken the harmonic structure factor to be the structure factor when the TO mode is fully recovered.}
\end{figure}

    The temperature dependences of the structure factors provide a sensitive measure of the coupling.  As the TO mode softens and the frequency gap between the TA and TO decreases for $q$ near the zone centre we expect the effects of coupling to be dramatic.  Specifically, as the two modes come closer together, a large transfer of spectral weight from one mode to another should occur in the presence of strong coupling.  The amount of transfer of spectral weight will characterize the strength of the coupling.  The structure factors are plotted in Fig. \ref{figure6} as a function of temperature and compared with those measured in SrTiO$_{3}$.  We have normalized all data by the structure factor measured when the TO mode is fully recovered.  In the case of PMN, the low-temperature structure factor at 100 K was used while for SrTiO$_{3}$ the data were divided by the high-temperature value.  This normalization puts the structure factors on the same scale and allows a direct comparison between different systems.

    The technique of characterizing the mode-coupling by tracking the integrated intensity has been used to investigate the dynamics in several materials.  The most relevant example is SrTiO$_{3}$ described in Ref. \onlinecite{Yamada69:26} which reports a significant anisotropy in the mode-coupling behavior based on structure factors.~\cite{Yamada69:26}  For acoustic T$_{2}$ modes, propagating along [1$\overline{1}$0], the integrated intensity was found to follow the prediction of harmonic theory.  In the case of T$_{1}$ modes, propagating along [100], the integrated intensity was measured to deviate significantly (by more than a factor of 20) from harmonic theory (illustrated by filled circles in Fig. \ref{figure6}).  Another example where structure factors were used to characterize the coupling is the superconducting borocarbides described in Ref.  \onlinecite{Bullock98:57} and Ref. \onlinecite{Stassis97:55}.  These works report a large transfer of spectral weight between the acoustic and optic branches, emphasizing the large coupling between the two modes. The temperature dependence of the structure factors in SrTiO$_{3}$ contrasts with that measured in PMN (open circles Fig. \ref{figure6}); the latter shows a comparatively small change in the measured structure factors with temperature (about a factor of two).

    For the T$_{1}$ mode measured near (200) we find exactly the same behavior as that measured for the T$_{2}$ mode.  The (200) structure factors map directly onto the intensity measured around (220) when corrected for the $Q^{2}$ factor in the harmonic formula for the energy integrated intensity.  We observe the same increase in spectral weight of the acoustic T$_{1}$ mode with decreasing temperature as well as little change in the acoustic frequency position.  We also observe similar symmetric lineshapes, well described by uncoupled modes, as presented elsewhere.~\cite{Wakimoto02:65, Wakimoto02:66}  Measurements similar to those described here have also been conducted in PZN near (220) and have found the same gradual increase in the acoustic integrated intensity below T$_{c}$.~\cite{Stock:unpub}  In comparison to SrTiO$_{3}$ and KTaO$_{3}$, the coupling between the acoustic and optic modes is very isotropic in PMN.

\section{Mode coupling and Comparison with the (110) Zone}

 \begin{figure}[t]
\includegraphics[width=8cm] {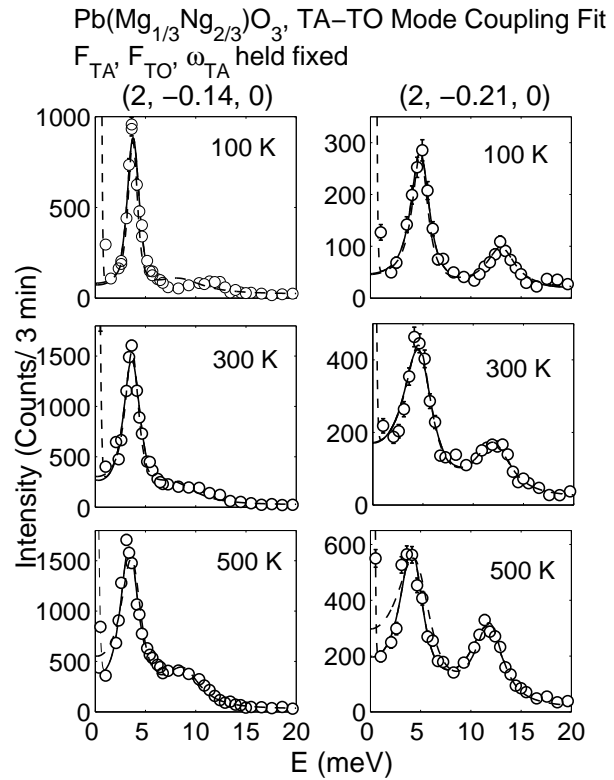}
\caption{\label{figure8} Results of mode coupling fits to data at (2.0, -0.14, 0) and (2.0, -0.21, 0) with the structure factors and coupling fixed as discussed in the text.  The dashed lines are fits to uncoupled modes ($\Delta=0$), but with temperature \textit{independent} structure factors}
\end{figure}

\begin{figure}[t]
\includegraphics[width=8cm] {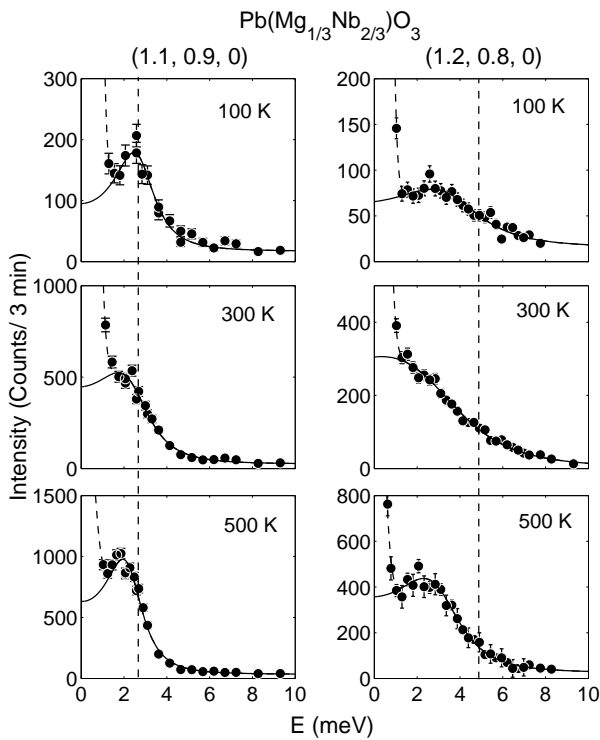}
\caption{\label{figure7} Constant-Q scans showing the temperature dependence of the T$_{2}$ mode near the (110) position.  An optic mode was not observable in our measurements indicating a small structure factor.  The vertically dashed lines indicate the peak position of the acoustic mode measured in the (220) zone.}
\end{figure}

    In PMN, the T$_{1}$ and T$_{2}$ acoustic modes are only weakly coupled to the optic modes as evidenced by the small change in the structure factors and the minimal shift in the acoustic frequency with temperature.  This contrasts with the behavior in SrTiO$_{3}$ where the T$_{2}$ mode shows little sign of coupling whereas the T$_{1}$ mode shows evidence for significant coupling.  The optic dispersion for both the T$_{1}$ and T$_{2}$ modes is also relatively isotropic.  Measurements near the (200) position yield $\hbar$$^{2}$$\alpha$=5.9 $\pm$ 1.0 $\times$ 10$^{2}$ meV$^{2}$$\cdot$\AA$^{2}$\ and near the (220) position $\hbar$$^{2}$$\alpha$=4.6 $\pm$ 0.2 $\times$ 10$^{2}$ meV$^{2}$$\cdot$\AA$^{2}$. Based on the symmetry of the dispersion and fact that the temperature dependence of the zone-centre TO mode measured near (200) and (220) agree exactly, we conclude that the TO mode is equally soft along the [100] and [110] directions.  This also contrasts with the large anisotropy in both the dispersion and temperature dependence observed in SrTiO$_{3}$ between the [110] and [100] directions.

    We do observe an anisotropy in the linewidth of the transverse acoustic mode.  For the acoustic T$_{1}$ mode we observe a broadening of the acoustic mode above T$_{c}$ whereas for the T$_{2}$ mode we observe no change in the linewidth around T$_{c}$.  We speculate that this anisotropy reflects the polarization direction of the correlated polar nanoregions at low-temperatures rather than anisotropy in the coupling.  We emphasize that the broadening observed for the acoustic T$_{1}$ mode is extremely small compared with that observed in the optic mode above T$_{c}$.~\cite{Wakimoto02:66}  We are currently working to extend this result to higher temperatures.

    To characterize the mode-coupling in terms of the lineshape in more detail, we have fit the formula described in Ref. \onlinecite{Currat89:40} and reproduced as Eqn. \ref{coupled} in the Appendix.  The coupled mode formula contains seven free parameters and therefore it is necessary to use other information to fix certain parameters.  We have fixed the structure factors to those values which give a good fit over a broad range in $q$=0.1-0.25 r.l.u. and temperature T=100-500 K near the (220) and (200) zones.  Based on this analysis we find that a value of $\mid F_{optic}/F_{acoustic}\mid^{2}$ $\sim$ 0.70 $\pm$ 0.15 describes the data reasonably well.  This value is comparable to that of 0.9, calculated based on the relative weight of Last and Slater modes discussed in Ref. \onlinecite{Hirota02:65}.

    Along with $q$ and temperature \textit{independent} structure factors (F$_{acoustic}$ and F$_{optic}$) a temperature independent coupling constant was used which varied with $q$ according to $\Delta=2.7 (\pm 0.3) \times 10^{2} \zeta^2-2.3(\pm 0.5)\times 10^3\zeta^4$ meV$^{2}$ (taking ${\bf{q}}$=$(\zeta, \zeta, 0)a^{*}$). The exact form of the coupling is defined and discussed further in the Appendix (see Eqn. \ref{coupled}).  This form for the real coupling was previously used to describe the coupling in KTaO$_{3}$.~\cite{Axe70:1}  We note that this $q$-dependent form for the coupling is exactly the low $q$ limit of that used in the calculations of Ref. \onlinecite{Hlinka03:91}.

    Based on the profiles shown in Fig. \ref{figure4} we do not expect the coupling to be strong as the phonon peaks are well described by symmetric Lorentzians.  Strong coupling has been previously observed to distort the phonon lineshapes dramatically due to phonon interference effects.  This effect has been observed in KNbO$_{3}$ in Ref. \onlinecite{Currat89:40} and in BaTiO$_{3}$ in Ref. \onlinecite{Harada71:4}.  In both these examples, a symmetric Lorentzian lineshape gives a poor description of the phonon profiles.

    The results of these fits are illustrated in Fig. \ref{figure8} for a series of temperatures at (2, -0.14, 0) and (2, -0.21, 0).  The agreement between the data and the model over a broad range of temperature and wave vector $q$ is excellent.  The dashed line in Fig. \ref{figure8} is the result of a fit to uncoupled oscillators with the structure factors fixed to the average value obtained from unconstrained fits at each temperature (shown in Fig. \ref{figure6}).  The agreement of the dashed line with the data is also very good showing only significant deviation at high temperatures.  This analysis illustrates further that the effects of coupling between the acoustic and optic modes are relatively small.

    After fixing the structure factors and coupling we still find that a temperature and wave vector dependent $\Gamma_{TO}$ (with $\Gamma_{TO}$ increasing with decreasing $q$) gives a good description of the data around T$_{c}$.  This implies that mode-coupling alone is inadequate to describe the anomalous damping of the TO mode in the ``waterfall'' region.  Instead, theories which involve coupling of the TO mode to some pseudospin (stochastic) variable~\cite{Yamada02:9573}, presumably generated by random fields~\cite{Stock03:xx,Westphal92:68}, may be more appropriate.  Such an extra coupling may capture the random nature of the polar nanoregions or any random fields introduced by chemical disorder.

    So far we have investigated the lattice dynamics in the (200) and (220) zones where the diffuse scattering has been shown to be weak or unobservable.  To investigate the influence of the diffuse component we have conducted constant-$Q$ scans in the (110) zone.  The (110) Brillouin zone is the simplest region in which to investigate the effects of the diffuse scattering as the optic structure factor is observed and calculated to be negligible.

    Fig. \ref{figure7} summarizes the results of the T$_{2}$ phonon near the (110) position, a zone where a strong temperature dependent diffuse signal has been observed.  In this zone we see a strong temperature dependence of the acoustic frequency position and lineshape, in contrast to that observed along the same direction in the (220) zone.  It is difficult to reconcile this behavior in terms of coupling between the acoustic and optic modes as the change in lineshape is much more dramatic than that measured in the (220) zone.  Furthermore, the optic structure factor has been measured to be extremely small. Therefore, its coupling should have no effect on the acoustic T$_{2}$ mode measured near (110).  Since the distortion of the TA mode in the (110) position cannot be attributed to the TO mode we speculate that there is a strong coupling between the diffuse component and the acoustic mode.  This has recently been further illustrated by the cold neutron work described in Ref. \onlinecite{Hiraka04:xx}.

    We speculate that it is this additional coupling to the central peak which is responsible for the anomalous lattice dynamics in Brillouin zones where a strong diffuse component is present.  All interpretations of the lattice dynamics have been based on a coupling between an acoustic and optic mode and these conclusions need to be revaluated.  We are currently working to extend our results to a (300) type zone where a strong optic and diffuse component exist.

    The importance of the diffuse component (or central) peak has recently been emphasized in Ref. \onlinecite{Gvasaliya04:69} in the context of interpreting relaxors in terms of order-disorder transitions.  Our results, suggest that the diffuse component also plays a much more important role in the lattice dynamics than any coupling between the acoustic and optic phonons as previously believed, and as is the case in conventional displacive ferroelectric systems like KTaO$_{3}$.  It is interesting to note that the apparent softening of the acoustic mode near (110) occurs close to T$_{c}$, and also in the same temperature region where the intensity and correlation length of the diffuse scattering have been observed to increase.~\cite{Xu03:xx}  This is strongly suggestive that the coupling between the central component and the phonons is important in the relaxor transition.

    Coupling between a vibrational mode and a central relaxational mode (diffuse component) has been observed in many systems and such a coupling can result in a soft mode analogous to a displacive transition.~\cite{Michel78:68,Bell94:66}  For example, such a model has been applied to calcite which displays both a soft mode and a continuum of scattering.~\cite{Harris98:10}  The strong correlation between the diffuse scattering and the acoustic mode in PMN may suggest that such models are suitable to explain the lattice dynamics behind the relaxor transition.  They, of course, follow naturally from the random field models of Refs. \onlinecite{Stock03:xx} and \onlinecite{Westphal92:68}.  We are currently working to extend our low-energy acoustic phonon work to other zones to characterize this coupling and its symmetry in more detail.

\section{Conclusions}

    The experiments presented here show that the lattice dynamics are very different between zones with a diffuse component and those where a central peak is absent.  This has been established by contrasting measurements made in the (220) and (200) zones with the (110) zone.  The effects of mode-coupling have been characterized by studying the lattice dynamics in the (200) and (220) zones and have been shown to be weak and isotropic compared with those in SrTiO$_{3}$. We also find that mode coupling cannot exclusively explain the anomalous damping of the TO mode observed in both the (200) and (220) zones.  Further work is currently being pursued to study the dynamics in relation to the diffuse scattering in other zones.

\begin{acknowledgements}

We are grateful to P.M. Gehring, H. Hiraka,  S. Wakimoto, and G. Xu for useful discussions and to A. Cull, L.E. McEwan, M.M. Potter,  R.Sabatini, and R. Sammon for invaluable technical assistance.  The work at the University of Toronto was supported by the Natural Science and Engineering Research Council and the National Research Council of Canada.  We also acknowledge financial support from the U.S. DOE under contract No. DE-AC02-98CH10886, and the Office of Naval Research under Grants No. N00014-02-1-0340, N00014-02-1-0126, and MURI N00014-01-1-0761.

\end{acknowledgements}

\section{Appendix: Phonon cross sections}

    In order to obtain detailed information on the linewidth from the inelastic spectrum, a model must be convolved with the resolution function and then fit to the observed scattering.  In this section we state the cross-sections used to describe the constant-Q data presented in the main part of the text.

     Measurement of the neutron scattering intensity provides a direct measure of S({\bf{q}}, $\omega$), which is related to the imaginary part of the susceptibility $\chi$$'$$'$($\bf{q}$,$\omega$) by the fluctuation dissipation theorem~\cite{Shirane:book},

\begin{eqnarray}
\label{optic}
S({\bf{q}}, \omega)=\pi^{-1}[n(\omega)+1]\chi''({\bf{q}},\omega),
\end{eqnarray}

\noindent where the energy transferred to the sample is defined by $\hbar \omega$=$E_{i}-E_{f}$ and $n(\omega$)=$1/(e^{\hbar \omega/k_{B}T}-1)$ is the Bose factor.  In the two sections contained in this Appendix we explicitly describe the two forms used for the uncoupled and coupled mode analysis.

\subsection{Uncoupled Harmonic Oscillators}

    To extract the integrated intensity (and hence the structure factor) as a function of temperature and wave vector we have described the phonon cross section for a particular mode by the antisymmetrized linear combination of two Lorentzians.  The imaginary part of the susceptibility is given by,

\begin{eqnarray}
\label{SHO} \chi''({\bf{q},}\omega)={A\over {[\Gamma(\omega)^{2}+
\left(\hbar \omega- \hbar \omega_{0}(\bf{q})\right)^{2}}]} \\
- {A\over {[\Gamma(\omega)^{2}+ \left(\hbar \omega+\hbar
\omega_{0}(\bf{q})\right)^{2}}]},
\nonumber
\end{eqnarray}

\noindent where $\Gamma(\omega)$ is the frequency dependent half-width-at-half-maximum (HWHM), $\omega_{0}(\bf{q})$ is the undamped phonon frequency, and A is the amplitude.  For acoustic modes, we have approximated the dispersion to be linear $\omega_{0}$=c$|{\bf{q}}|$.  For optic modes, we have set the dispersion to have the form described in Eqn. \ref{optic}.

\subsection{Coupled-Mode Analysis}

    To study the effects of mode coupling on the lineshape we have used the coupled mode formula previously derived and used to study perovskite ferroelectric systems.~\cite{Currat89:40,Wehner75:36}  For a pair of interacting modes, the imaginary part of the susceptibility can be written as

\begin{eqnarray}
\label{coupled} \chi''({\bf{q},}\omega)={(C Y-B Z)\over(B^{2}+C^{2})},
\end{eqnarray}

\noindent where,

\begin{eqnarray}
\nonumber
B=(\omega_{1}^{2}-\omega^{2})(\omega_{2}^{2}-\omega^{2})-\omega^{2}\Gamma_{1}\Gamma_{2}-\omega_{1}\omega_{2}\Delta^{2},
\end{eqnarray}
\begin{eqnarray}
\nonumber
C=\omega[\Gamma_{1}(\omega_{2}^{2}-\omega^{2})+\Gamma_{2}(\omega_{1}^{2}-\omega^{2})],
\end{eqnarray}
\begin{eqnarray}
\nonumber
Y=F_{1}^{2}(\omega_{2}^{2}-\omega^{2})+F_{2}^{2}(\omega_{1}^{2}-\omega^{2})-2F_{1}F_{2}(\omega_{1}\omega_{2})^{1/2}\Delta,
\end{eqnarray}
\begin{eqnarray}
\nonumber
Z=\omega(F_{1}^{2}\Gamma_{2}+F_{2}^{2}\Gamma_{1}).
\end{eqnarray}

\noindent The frequency positions and dampening constants for the two interacting modes are given by $\omega_{1}$, $\omega_{2}$, $\Gamma_{1}$, and $\Gamma_{2}$ respectively.  The structure factors for each mode are given by F$_{1}$ and F$_{2}$.  The real coupling constant is defined by $\Delta$.  In our analysis we have followed previous works and assumed the coupling constant to be real.  A generalized formula with both imaginary and real coupling constants can be found in Ref. \onlinecite{Currat89:40}.

    It is clear from the formula above that, in the presence of coupling, a temperature dependent frequency will result in an apparent temperature dependence (as measured from the integrated intensity of the two phonon peaks in the intensity profile) in the structure factors.  This was illustrated in the early work of Ref. \onlinecite{Axe70:1} and provides the basis for our fits over a broad range in temperature and energy transfer to the phonons in the (200) and (220) zones.


\end{document}